\documentstyle[12pt]{article}
\newcommand{\be}{\begin{equation}}
\newcommand{\ee}{\end{equation}}
\begin{document}
\begin{titlepage}
\title{Big smash of the universe}
\author{Valerio Faraoni\\ \\
{\small \it Physics Department, University of Northern British Columbia}\\
{\small \it 3333 University Way, Prince George, B.C., Canada V2N~4Z9}\\
{\small \it email~~vfaraoni@unbc.ca} 
}
\date{} 
\maketitle   \thispagestyle{empty}  \vspace*{1truecm}
\begin{abstract} 
The occurrence of a big smash singularity which ends the universe in a
finite time in the future is investigated in the context
of superquintessence, i.e. dark energy with effective equation of state
parameter $ w< -1 $ and $\dot{H}>0$. The simplest toy model of
superquintessence
based on a single nonminimally coupled  scalar field exhibits big smash
solutions which are attractors in phase space.
\end{abstract}
\vspace*{1truecm} 
\begin{center} 

\end{center}     
\end{titlepage}  


A new picture of the universe has emerged in recent years: the data from
the {\em Boomerang}
\cite{Boomerang}
and {\em MAXIMA} \cite{Hananyetal00} experiments confirm
that we live in a  spatially flat ($\Omega=1$) universe, where $\Omega
=\Omega^{(m)} + \Omega^{(q)} $ is
the total energy density expressed in units of the critical density
\cite{footnote}.
Baryonic and dark matter only account for $\Omega^{(m)} \simeq
0.3$ of the total energy density $\Omega$, while the rest is due to a yet
unknown form of {\em dark energy}. Studies of Type Ia supernovae \cite{SN}
and of radio-galaxies \cite{Dalyetal} show that the
present expansion of the universe is
accelerated, i.e. $\ddot{a} >0$, where $a(t)$ is the scale factor of the
Friedmann-Robertson-Walker line element 
\be   \label{FRWmetric}
ds^2=-dt^2 +a^2 (t) \left( dx^2 + dy^2 + dz^2 \right)  
\ee
describing our universe in comoving coordinates $\left( t,x,y,z \right)$. 
The Friedmann equation 
\be
\frac{ \ddot{a} }{a} =-\, \frac{\kappa}{6} \, \left( \rho+3P \right) \;,
\ee
where $\rho $ and $P$ are, respectively, the energy density and pressure
of the source of gravity, shows that in order to have accelerated
expansion the pressure of the dark energy dominating the dynamics
must be negative, $P< -\rho/3$. 
A cosmological
constant $\Lambda$ as the explanation of dark energy is rejected by most
cosmologists because of the cosmological
constant problem \cite{Weinberg89} and of the cosmic coincidence problem.
The vacuum energy density predicted
by
high energy physics with a  Planck scale cutoff is wrong by 120 orders of
magnitude (or 40 orders of magnitude if the cutoff is at the QCD  
scale). Solving the
{\em cosmic coincidence problem} of why the dark energy came to
dominate the dynamics only recently (at redshift $z\sim 1$) requires
extreme fine tuning of $\Lambda$. One would rather have $\Lambda $ exactly 
equal to zero due to some yet unknown mechanism than this extreme
fine-tuning, and it is preferable to explain the present cosmic
acceleration with a dynamical
vacuum energy called {\em quintessence}. Many models of quintessence, 
most of which based on scalar fields, have been proposed.

Observational efforts aim at
determining the effective equation of state parameter of the universe $w
\equiv P/\rho$. It was pointed out that the current data allow for
\cite{HannestadMortsell02,Dalyetal,Melchiorrietal02,Schueckeretal02},
or even favour
\cite{Caldwell99}-
\cite{Carrolletal03},
values of this parameter in the $w< -1 $
range. Were a value $w<-1 $ to be confirmed by observations, it would be a
most interesting finding, because such values cannot be explained by
Einstein gravity with  a minimally coupled scalar, if one assumes
positivity of the energy density. In
fact in such ``canonical'' models the scalar field has energy density
and pressure 
\be   \label{mcdensity}   
\rho=\frac{ \left( \dot{\phi} \right)^2 }{2}+V( \phi) \;, \;\;\;\;\;\;\;\;
P =\frac{ \left( \dot{\phi} \right)^2 }{2} -V( \phi) \;,
\ee
and effective equation of state parameter
\be
w=\frac{x-1}{x+1} \;,
\ee
where $x \equiv \left( \dot{\phi} \right)^2/ \left( 2V \right)  \geq 0 $
if
$V > 0$, giving
$-1 \leq w \leq 1$ (the minimum of $w$ being attained by de Sitter
solutions). In these models one only obtains $w < -1$ by assuming $V<
-\left( \dot{\phi} \right)^2 /2 \leq 0 $,  which in turn implies a
negative
energy density (\ref{mcdensity}), and $\rho <0$ is 
hardly an acceptable proposition. Correspondingly, the
Friedmann equation  
\be
\dot{H}=-\, \frac{\kappa}{2} \left( \rho + P \right) 
\ee
tells us that, for a minimally coupled scalar in Einstein's gravity 
it is $\dot{H} = -
\kappa \left( \dot{\phi} \right)^2 /2 \leq 0$ (the
extreme case $\dot{H}=0$ again corresponding to de Sitter
solutions). A
regime with $w<-1$ is associated to $\dot{H}>0$ and is called {\em
superacceleration}; a form of dark energy capable of sustaining
superacceleration was dubbed {\em superquintessence} \cite{Faraoni02} or
{\em phantom energy} \cite{Caldwell99}.

Models have been proposed to explain superacceleration regimes, including
scalar fields nonminimally coupled to the Ricci curvature, actions with
the ``wrong'' sign of the
kinetic  energy of the scalar, supergravity-inspired models with
non-canonical kinetic energy terms and zero potential ({\em
k-essence}), Brans-Dicke-like fields in
scalar-tensor gravity, or stringy matter \cite{models}. If the universe
superaccelerates, its expansion becomes 
so fast that it risks ending its existence in a finite time, $a(t)
\rightarrow \infty $ as $t\rightarrow t_0$ with $t_0$ finite. For
solutions
with this property ({\em big smash} or {\em big rip} solutions)
\cite{Caldwell99,Starobinsky00,McInnes02,FramptonTakahashi02,
Caldwelletal03}
the energy
density of superquintessence {\em increases} with time instead of
redshifting away as the matter or radiation energy densities $\rho^{(m)}
\propto a^{-3}$, $\rho^{(r)} \propto a^{-4}$, or as the
energy density of ``ordinary'' quintessence. Big smash solutions occur in
the theory of ordinary 
differential equations (ODEs) when the solutions of an ODE 
cannot be maximally extended to an infinite interval. An example of ODE
which does not admit maximal extension of its solutions is
\be  \label{exampleODE}
\frac{d y(x)}{dx}= A\, y^2 (x)  \; ;
\ee
if $A>0$ the rate of change of the solution is fed by the
increasing value of $y$ itself in a positive feedback mechanism
that makes $y(x)$ grow so fast that it explodes in
a finite time, while this behaviour is absent if~$ A  \leq 0 $. 
As we shall see, eq.~(\ref{exampleODE}) is similar to the equation
satisfied by the Hubble parameter of a superaccelerating universe.

A big smash can  be avoided in certain models of
superquintessence (e.g. \cite{Ziaeepour}) but is a generic feature in
other models
\cite{Caldwell99}. Whether the big smash is
unavoidable or not depends, of
course, on the model adopted \cite{footnote2}: here we consider what is
perhaps the
simplest model of
superquintessence, namely a single
scalar field $\phi$ coupled nonminimally to the Ricci curvature, described
by the action
\be  
S =  \int d^4 x \sqrt{-g} \left[  \frac{R}{2} \left( \frac{1}{\kappa}
-\xi \phi^2 \right)  -  \frac{1}{2}  g^{cd} \, \nabla_c \phi \,
\nabla_d \phi  - V( \phi ) \right]  + S^{(m)} \; ,
\label{action}
\ee 
where $\xi $ is a dimensionless coupling constant and $S^{(m)} $ is the
action for ordinary matter. Nonminimal coupling is introduced by
renormalization even if it is absent at the classical level and is
required in classical general relativity by the Einstein equivalence
principle \cite{SonegoFaraoni93}. The field equations are 
\be \label{Gab}
G_{ab}=\kappa \, T_{ab} \left[ \phi \right] \;,
\ee
\be 
T_{ab} \left[ \phi \right]  =\nabla_a \phi \, \nabla_b \phi
-\frac{1}{2} \,
g_{ab} \nabla^c 
\phi \, \nabla_c \phi -V \, g_{ab}  
+ \xi \left( g_{ab} \Box
-\nabla_a 
\nabla_b \right) \left( \phi^2 \right) +\xi G_{ab} \phi^2 \; . 
\label{Tab}
\ee
The gravitational coupling in eq.~(\ref{Gab}) is the usual 
constant $\kappa=8\pi G$ and not the effective time-dependent coupling
$\kappa_{eff} = \kappa \left( 1-\kappa \xi \phi^2 \right)^{-1}$ recurring 
in the literature and corresponding to a different way of writing the
field equations \cite{BellucciFaraoni}. 
Moreover, the scalar field stress-energy tensor (\ref{Tab}) (``improved
energy-momentum tensor'' \cite{CCJ70}) is covariantly conserved, $\nabla^b
\,
T_{ab}[ \phi ] =0$. In the metric (\ref{FRWmetric}), the field equations
become
\be 
 6\left[ 1 -\xi \left( 1- 6\xi \right) \kappa \phi^2
\right] \left( \dot{H} +2H^2 \right)  
 -\kappa \left( 6\xi -1 \right) \dot{\phi}^2   
- 4 \kappa V  
+ 6\kappa \xi \phi \, \frac{dV}{d\phi} = 0 \; ,
\label{fe1} 
\ee
\be
\frac{\kappa}{2}\,\dot{\phi}^2 + 6\xi\kappa H\phi\dot{\phi}
- 3H^2 \left( 1-\kappa \xi \phi^2 \right) + \kappa  V =0  \, ,
 \label{fe2} 
\ee
\be \label{KG}
\ddot{\phi}+3H\dot{\phi}+\xi R \phi +\frac{dV}{d\phi} =0 \; .
\ee
Only two of these equations are independent, as the Klein-Gordon
equation (\ref{KG}) can be derived from the conservation equation $\dot{ {
\rho^{(\phi)} } } + 3H
\left( P^{(\phi)} +\rho^{(\phi)} \right)=0$ when $\dot{\phi} \neq 0$,
where 
\be \label{15}
\rho^{(\phi)} = \frac{\dot{\phi}^2}{2} + V(  \phi)
+3\xi H\phi \left( H\phi+ 2 \dot{\phi}\right)
\ee
and
\be 
P^{(\phi)} =  \frac{\dot{\phi}^2}{2}-V( \phi ) -\xi \left[ 4H\phi
\dot{\phi}  +2\dot{\phi}^2 +2\phi \ddot{\phi} 
 + \left( 2\dot{H}+3H^2 \right) \phi^2 
\right]  \label{18bis}
\ee
are, respectively, the effective energy density and pressure of the
fluid equivalent to the nonminimally
coupled scalar. Note that the Hamiltonian constraint (\ref{fe2}) can be
written as
\be 
H^2=\frac{\kappa}{3} \, \rho^{(\phi)}  \;,
\ee
which is consistent with $\rho^{(\phi)} \geq 0$. This is not the case
with
other 
definitions of effective  energy-momentum tensors $T_{ab}[ \phi ]$ used in
the literature (see the discussion in Ref.~\cite{BellucciFaraoni}). 
In order to investigate the fate of the universe with respect to big
smash solutions, we neglect the matter part of the action
$S^{(m)}$; this assumption is justified by the fact that, when
a superacceleration regime sets in, the superquintessence energy density
$\rho^{( \phi) } $ quickly grows to dominate the matter energy density,
which instead fades away.

Only the two variables $H$ and $\phi$ are needed to describe the dynamics
of the system (\ref{fe1})-(\ref{KG}), and the phase space is
a two-dimensional manifold with  a rather complex
structure \cite{Gunzigetal}. This
is best seen by rewriting the field equations as
\cite{Gunzigetal} 
\be   \label{phidot}
\dot{\phi}=-6\xi H \phi \pm \frac{1}{2\kappa} \, \sqrt{ {\cal F} \left( H,
\phi
\right)} \;,
\ee
\begin{eqnarray}   \label{hdot}
\dot{H} = \left[  
3\left( 2\xi-1 \right) H^2 +3\xi  \left( 6\xi -1 \right) \left( 4\xi-1
\right) \kappa H^2 \phi^2 
\mp  \xi \left( 6\xi-1\right) \sqrt{
{\cal F}} \, H\phi  \right. \nonumber \\  
\left. + \left(
1-2\xi \right) \kappa V   -\kappa \xi \phi \,
\frac{dV}{d\phi} \, \right]
\, \frac{1}{
1+\kappa\xi \left( 6\xi-1 \right) \phi^2} \;,
\end{eqnarray}
where 
\be
{\cal F} \left( H, \phi \right) = 8\kappa^2 \left[ \frac{3H^2}{\kappa} -V(
\phi )
+3\xi \left( 6\xi -1 \right) H^2 \phi^2 \right] \;.
\ee
The appearance of the $\pm $ signs in eqs.~(\ref{phidot}) and (\ref{hdot})
requires a
clarification: the phase space curved manifold is composed of two sheets,
corresponding to upper and lower sign, 
and there  typically  is a region forbidden to the dynamics
corresponding
to $ {\cal F} \left( H, \phi \right) <0$; the two sheets join each other
at the
boundary of this forbidden region, which needs not be simply connected.
Orbits of solutions lying in the ``upper''  sheet can switch to the  
``lower'' sheet at
these points, and vice-versa \cite{Gunzigetal}. Far
from the forbidden region's boundary, the orbit of a solution is forced to
stay in one sheet and cannot cross over to the other sheet without
approaching the forbidden region and touching its boundary. Hence, for
large values of $H$ and
$\phi$ only one sign in
eqs.~(\ref{phidot}) and (\ref{hdot}) applies to each solution.  Since
$H(t)$ and $\phi(t)$
grow so quickly when the superacceleration regime sets in, it is
meaningful to perform an asymptotic analysis for large values of these
variables when searching for big
smash solutions.  We consider conformal coupling $\xi=1/6$, which is a
stable infrared fixed point of the renormalization group
\cite{onesixth}, and consider as a toy model the
potential 
\be  \label{potential}
V( \phi ) =\frac{m^2 \phi^2}{ 2} + \lambda \phi^4 \;,
\ee
where $\lambda$ will be required to be negative. This does
not harm the positivity of the energy density $ \rho^{ (\phi)} $ when
$\xi \neq 0$.  Eqs.~(\ref{phidot}) and
(\ref{hdot}) then reduce to
\be  \label{approxphidot}
\dot{\phi} \simeq - H \phi \pm \sqrt{-2\lambda} \, \phi^2 \;,
\ee
\be  \label{approxhdot}
  \dot{H} \simeq  -2 H^2 +\mu \phi^2 \;,
\ee
where $ \mu \equiv \kappa m^2/6 $. We look for big smash solutions of the
form
\be  \label{trial1}
a(t) = \frac{ a_* }{  \left| t- t_0 \right|^{\alpha_{\pm} } }  \;,
\ee
and
\be  \label{trial2}
\phi(t)= \frac{ \phi_*}{ \left| t-t_0 \right|^{\beta_{\pm} } } \;,
\ee
with $ \alpha_{\pm}, \beta_{\pm} >0 $, consistently with the
approximation of large $H$ and $\phi$ employed, and where $t_0$,  $a_* $
and
$\phi_* $ are constants; $t_0$ will be approached from below. Substitution
of eqs.~(\ref{trial1}) and (\ref{trial2}) into eqs.~(\ref{approxphidot})
and (\ref{approxhdot}) yields
\be  \label{alpha}
 \alpha_{\pm} = \frac{ \pm \sqrt{ -\lambda \left( 2\mu +\lambda \right) }
\,
-\left(
\mu+\lambda \right) }{ \mu+4\lambda } \;,
\ee
\be
\beta_{\pm}=1 \;,
\ee
\be   \label{phistar}
\phi_{*}^{\pm}= \pm \,\, \frac{ 1+\alpha_{\pm} }{ \sqrt{-2\lambda} } \;.
\ee
By taking the positive sign in eq.~(\ref{alpha}), one immediately sees 
that there are big smash solutions ($\alpha >0$) in the range of
parameters 
\be
1 < \frac{ \mu }{ \left| \lambda \right| } < 4 \;.
\ee
Next, one would like to know whether these big smash solutions are
stable or if they disappear when perturbed. The equations for
the perturbations $\delta H$ and $ \delta \phi$ are sufficiently involved
to
defy a direct analytical investigation. It is convenient, instead, to
focus on the projection onto the $\left( H, \phi \right) $ plane of the
two sheets composing the phase
space. Assuming $\dot{H}, \dot{\phi} \neq
0$, as is legitimate during the late stages of a superacceleration regime,
one obtains the
vector field
\be
\frac{dH}{d\phi} =\frac{\dot{H}}{ \dot{\phi} } = \frac{2u^2-\mu}{u \mp
\sqrt{-2\lambda} } \;,
\ee
where $u \equiv H/\phi$. The identity
\be
\frac{dH}{d\phi}= u+\phi\, \frac{du}{d\phi} 
\ee
then yields
\be  \label{eqforu}
\frac{du}{d\phi}= \frac{ u^2 \pm \sqrt{-2\lambda} \, u -\mu}{ u \mp
\sqrt{-2\lambda} } \, \left(  \frac{1}{\phi} \right) \;.
\ee
It is straightforward to see that
exact solutions of  eq.~(\ref{eqforu}) with $u=$const.$\neq \pm
\sqrt{-2\lambda}$ exist and are given by
\be  \label{solutionshphi}
H = \frac{ \mp \sqrt{-2\lambda} \, \pm \sqrt{ -2\lambda+4\mu} }{2} \,
\, \phi \;.
\ee
Eq.~(\ref{solutionshphi}) includes the big smash solutions
(\ref{trial1})-(\ref{phistar}). In fact, simple algebra shows that
for the latter
\be
u = \frac{\alpha_{\pm}}{\phi_*} =
- \,  \frac{   \sqrt{ 4\mu -2\lambda }\, +\sqrt{-2\lambda}  }{2} \;,
\ee
which reproduces a special case of eq.~(\ref{solutionshphi}).

Let us proceed to study the stability with respect to linear
perturbations of the solutions of eq.~(\ref{eqforu}), which can be
rewritten as  
\be
H=\gamma_i \, \phi \;,
\ee
where the constant $\gamma_i$ can assume the values
\begin{eqnarray}
\gamma_1 & =&  \frac{ -\sqrt{-2\lambda}\,+ \sqrt{ 4\mu -2\lambda }\,
}{2} >0 \;, \\
\gamma_2 & =& 
\frac{ -\sqrt{-2\lambda}\,- \sqrt{ 4\mu -2\lambda }\,   }{2} < 0 \;, \\
\gamma_3 & =& -\gamma_1 < 0
\;, \\
\gamma_4 & =&   -\gamma_2 > 0
\end{eqnarray}
(the big smash solutions (\ref{trial1})-(\ref{phistar}) corresponding to
$\gamma_2$). The perturbations $\delta u$ in 
\be \label{uperturbations}
u( \phi) =u_0 +\delta u =\gamma_i +\delta u \left( \phi \right) \;,
\ee
satisfy the equation
\be
\frac{d \left( \delta u \right)}{d\phi} =\frac{
2u_0 \, \delta u \pm \sqrt{-2\lambda} }{ u_0 \mp \sqrt{-2\lambda} } \,
\frac{\delta u}{\phi} 
\ee
which yields 
\be
\delta u = \epsilon \, \phi^{\delta_i} \;,
\ee
where $\epsilon$ is a constant and 
\be \label{deltai}
\delta_i= \frac{ 2\gamma_i \pm \sqrt{-2\lambda} }{\gamma_i \mp
\sqrt{-2\lambda} } \;.
\ee
There are the following possibilities for $ \delta_i$, corresponding to
the four values of $\gamma_i$ and to the upper and lower sign in
eq.~(\ref{deltai}):
\be 
\delta_{1a}= \frac{ 2\sqrt{4\mu-\lambda}}{
-3\sqrt{-2\lambda}+\sqrt{4\mu-\lambda}} \;,
\ee
which is positive if $\mu/\left| \lambda \right| > 17/12 \simeq 1.4167$;
this range of values of the parameters corresponds to growing
perturbations and instability, while the finite range $ 0 < \mu/\left|
\lambda
\right| < 17/12$ corresponds to stability.
\be 
\delta_{1b}= 2\, \frac{ \sqrt{4\mu-\lambda} -2\sqrt{-2\lambda}   }{
\sqrt{-2\lambda}+\sqrt{4\mu-\lambda}} 
\ee
is positive if $\mu/\left| \lambda \right| > 7/24\simeq 0.2917$;
this range corresponds to instability.
\be
\delta_{2a}= \frac{ 2\sqrt{
4\mu-\lambda}}{-2\sqrt{-2\lambda} -\sqrt{4\mu-\lambda}} < 0 
\ee
corresponds to the big smash solutions and to stability for any
value of the parameters $\lambda<0$ and $\mu$; on the other hand, the
possibility
\be
\delta_{2b}= 2 \, \frac{ \sqrt{4\mu-\lambda}
+2\sqrt{-2\lambda} }{ \sqrt{4\mu- \lambda} - \sqrt{-2\lambda} } >0 
\ee
corresponds to instability. The remaining cases give the values of
$\delta$ 
\begin{eqnarray}
\delta_{3a} &= & \delta_{1b} \;, \;\;\;\;\;\;\;\;\;
\delta_{3b}=\delta_{1a} \; , \nonumber \\
\delta_{4a} & = & \delta_{2b} \;, \;\;\;\;\;\;\;\;\;
\delta_{4b}=\delta_{2a}
\end{eqnarray}
already considered.

The big smash solutions are stable against linear
perturbations and behave as attractors in phase space. Thus, there is a
finite
chance
that a universe described by the model considered here end its existence
in a finite time due to a big smash with infinite expansion, in which the
energy density diverges instead of being diluted away and bound
systems are gradually ripped apart \cite{Caldwelletal03}.

A few considerations are in order: first, the value of the effective
equation of state parameter $w$ is still subject to uncertainty and a
value $w<-1$ associated to superacceleration is not yet
confirmed; second, even if such a value were supported by future
experiments, it does not automatically imply that the universe will end in
a big smash. Third, the toy model of superquintessence employed here
should be generalized to more realistic models, which however are not
constrained sufficiently well by the presently available data. Potentials 
used to model quintessence are usually different from
(\ref{potential}), but the latter is
a very common potential in scalar field cosmology and is easier to study
as a
toy model (it is difficult to reach conclusions about the existence of
big smash solutions with other potentials, and even more difficult to
perform a stability analysis). The presence
and stability of big smash solutions in a finite future for a wide range
of parameters leads one to regard
a big smash as a  generic feature of scalar field models of
superquintessence that include a nonminimal coupling to gravity. This can
be of the simple form described by the action (\ref{action}), or of a more
general form as in scalar-tensor theories, which have been known to
contain smash solutions for  a long time \cite{STsmash}. The action
(\ref{action}) can be
explicitly reformulated as a scalar-tensor theory with variable
Brans-Dicke parameter 
\be
\omega (\varphi) = \frac{G\varphi}{4\xi \left( 1-G\varphi \right)} \;,
\ee
and
\be
\varphi = \frac{1- \kappa \xi \phi^2}{ G } \;,
\ee
but more general forms of
the coupling function $\omega ( \phi) $ are possible.  Since the field
equations
for the coupled variables $H$ and $\phi$ in scalar-tensor gravity
exhibit terms
similar to the right hand side of eq.~(\ref{exampleODE}),
solutions with explosive growth are possible in these
theories.

The most stringent constraint on the theory of the nonminimally coupled
scalar comes from Solar System experiments. The Brans-Dicke-like field
$\varphi$ mediates a long range force that is constrained by tests of
general relativity. Since $\varphi$ varies on a cosmological time scale,
it is appropriate to approximate $ \varphi$ with its present value
$\varphi_0$ and $ \omega \left( \varphi \right) \simeq 
\omega \left( \varphi_0 \right) \equiv \omega_0 $. The lower bound on
$\omega_0$ is\footnote{The more stringent bound $ \left| \omega_0 \right|>
3300 $ \cite{Will98} yields a constraint on  $ \left| \xi \right| $ of the
same order of magnitude.}  $ \left|
\omega_0 \right| > 500 $ \cite{Reasenbergetal79,Will93}, which yields
\be
\left| \omega_0 \right| =\left| \frac{ 1-\kappa \, \xi \phi_0^2 }{ 4\kappa
\, \xi \phi_0^2 }  \right|> 500 \;.
\ee
Although the present day value of $\phi$ is unknown, a weak coupling
regime in which $\kappa \, \left| \xi \right| \phi_0^2 << 1$ is plausible
\cite{Chiba}, given that typical values of $\xi$ predicted by
renormalization are of the order of $10^{-1}$-$10^{-2}$, and that $\phi_0$
cannot exceed the Planck mass $m_{pl}=G^{-1/2} $ by too much without
causing fine-tuning problems in the parameters $m$ and $\lambda$ (the
energy scale $V$ must be below the Planck energy scale $m_{pl}^4$). In the
weak coupling
regime $\kappa \left| \xi \right| \phi_0^2 << 1$ and assuming that $
\sqrt{\kappa} \, \left| \phi_0 \right| \simeq 1 $ (corresponding to $
\phi_0
\simeq 0.2 m_{pl}$), one obtains $ \left| \xi \right|< 5\cdot 10^{-4}$
(this limit is weakened if $\left| \phi \right| << m_{pl}$). This 
constraint limits the amount of superacceleration that is present today
(see e.g. Ref.~\cite{Torres}), but it should be kept in mind that even a
small amount of superacceleration can eventually lead to a big smash.
Stringent limits on $\left| \xi \right|$ can be satisfied and the big
smash will occur later: whether this amounts to fine-tuning the value of
$\xi$ is determined by the still unknown value of the parameter $w$
which quantifies the present amount of superacceleration
(assuming that the universe really does superaccelerate, i.e. that
$w<-1$). It is hoped that the observational determination of the value of
$w$ will soon clarify this issue.

To conclude, even if the departures of gravity from general relativity are
small today in the Solar System, on a large scale they may have a
catastrophic effect on the future of the universe. Usually, research 
on scalar-tensor cosmology has focussed on how scalar-tensor theories can
depart from Einstein's gravity in the early universe and converge to it at
later epochs. Here, the issue is rather the one  of a cosmological
solution of scalar-tensor gravity that is close to a general relativistic
solution today (when superacceleration is still moderate and $w$ close to 
$-1$), but will dramatically depart from it
in the future. If the universe really superaccelerates, the concern about
a big smash is legitimate. It is intriguing that observational data place
our present universe so close to the boundary $w=-1$ between the
possibility of  a big smash and certain evolution into infinite
dilution in an infinite time.

\vskip0.5truecm
\noindent {\small The author acknowledges Leon Brenig for a
stimulating discussion at a meeting in Peyresq.}

   
\end{document}